\documentstyle[psfig,conf_iap]{article}

\def\deg{$^\circ$}
\def\kms{km s$^{-1}$~}
\def\HII{\hbox{H\,{\sc ii}}}
\def\CII{\hbox{C\,{\sc ii}}}
\def\HI{\hbox{H\,{\sc i}}}
\def\CO{$^{12}$CO}
\def\lv{\hbox{\em lv}}
\def\arcmin{$^\prime$}

\begin{document}

\heading{Low-frequency Carbon Recombination Lines towards \\ \HI\ self-absorption
features in the Galactic Plane }

\par\medskip\noindent
 
\author{D. Anish Roshi$^{1,2}$, Nimisha G. Kantharia$^{2}$, K. R. Anantharamaiah$^{3}$} 

\address{National Radio Astronomy Observatory, Green Bank, WV 24944, USA} 
\address{National Centre for Radio Astrophysics, TIFR, Pune 411007, India}
\address{Raman Research Institute, Bangalore 560 080, India}

\begin{abstract}
A survey of radio recombination lines (RLs) in the Galactic
plane ($l = $ 332\deg $\rightarrow$ 89\deg)
near 327 MHz made using the Ooty Radio Telescope (ORT) 
has detected carbon RLs from all the positions 
in the longitude range 0\deg $< l < $ 20\deg\ and from a few positions at 
other longitudes. The carbon RLs detected in this survey originate from
``diffuse'' \CII\ regions. Comparison of the \lv\ diagram and 
the radial distribution of carbon line emission  
with those obtained from hydrogen RLs near 3 cm from \HII\ regions and ``intense'' \CO\ emission
indicates that the distribution of the diffuse \CII\ regions in the inner
Galaxy is similar to that of spiral arm components. 
Towards several observed positions, the central velocities of
the carbon RLs coincide with \HI\ self-absorption
features suggesting an association of diffuse \CII\ regions with 
cool \HI.  The observed widths of the lines from the two species are,
however, different. We discuss possible reasons for the 
difference in the line widths.
\end{abstract}

\section{Introduction}

Recombination lines of carbon originate either in regions adjacent to
\HII\ regions (classical \CII\ regions) or in neutral components (\HI\ or molecular) 
of the interstellar
medium (diffuse \CII\ regions).  
The diffuse \CII\ regions are identified through  observations of carbon RLs 
in absorption at frequencies below $\sim$ 100 MHz
and in emission above $\sim$ 150 MHz \cite{ka01}. 
This paper concerns new extensive
observations of the diffuse \CII\ regions in carbon RLs  at
frequencies near 327 MHz. 

\section{Observations and Results}

RL observations were made in the
longitude range $l =$ 332\deg $\rightarrow$ 89\deg, 
using the ORT in two different angular resolutions
--  2\deg $\times$ 2\deg\ and  2\deg $\times$ 6\arcmin \cite{ra01}.
In the low resolution survey, carbon RLs were detected
at almost all the positions in the longitude range $l = $ 0\deg\ to
20\deg\ and at a few positions between $l = $ 20\deg\ to 89\deg.

The \lv\ diagram and the radial
distribution, obtained from carbon RL emission near 327 MHz, show
similarity with those obtained from hydrogen RLs observed near 3 cm from \HII\ regions
and ``intense'' \CO\ emission. This similarity indicates that 
the distribution of the diffuse \CII\ regions in the inner Galaxy is 
similar to that of the spiral arm components.

\section{Association with cool \HI\ gas}

\begin{figure}
\centerline{\vbox{
\psfig{figure=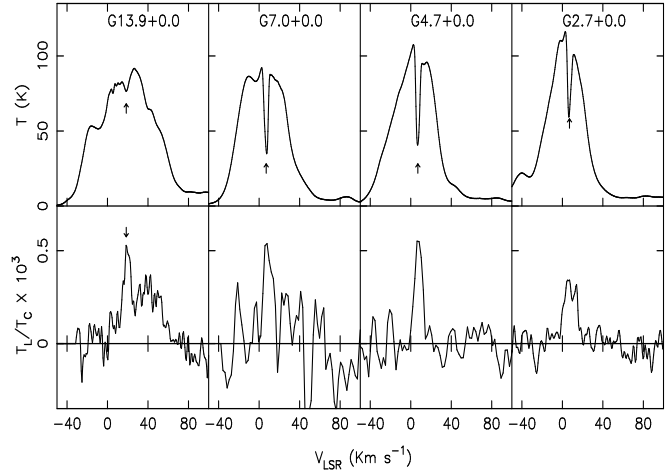,width=8.cm,height=4.8cm,rwidth=8.cm,rheight=4.8cm}
}}
\caption[]{Carbon RLs near 327 MHz (bottom; angular resolution 2\deg $\times$ 2\deg) and 
\HI\ spectra (top; from \cite{hb97}), obtained after convolving the data to 2\deg\ $\times$ 2\deg\ resolution, towards the
positions marked on each frame. 
The coincidence of the central velocities of the spectral features of the two species are clearly seen.
Towards G13.9+0.0 only the narrow carbon RL component shows velocity coincidence with the indicated \HI\ self-absorption
feature}
\end{figure}

Towards several positions in the inner Galaxy, the central velocities of
the observed (2\deg\ $\times$ 2\deg\ resolution) carbon RLs coincide with known \HI\ self-absorption
features. \HI\ self-absorption is observed when a cool ($\sim$ 20 K) \HI\
cloud appears in absorption against \HI\ emission from warmer gas behind
it. The similarity of the central velocities suggests that the diffuse \CII\ regions
in these directions coexist with the cool \HI\ clouds. The
observed widths ($\sim$ 4 \kms) of the HI self-absorption features are,
however, much narrower than the typical widths ($\sim$ 15 \kms) of the
carbon RLs. The difference in line widths may be due to :
(1) There may be several cool \HI\ clouds with slightly different central
velocities within the 2\deg $\times$ 2\deg\ beam and most of them may produce
carbon lines, which results in a broader RL. 
%If there are several cool \HI\ clouds with different central
%velocities within the beam used for RL observation and if most of these clouds emit
%carbon lines, then the recombination line observations will detect a
%broader carbon line. 
However, the \HI\ observations will detect only the
coldest HI clouds and hence the line will have a narrow line width.
(2) Alternatively, if ``warmer'' ($\sim$ 100 K), presumably more turbulent \HI\ gas 
is associated with cool \HI\ clouds then both ``warm'' and ``cool'' gas will emit
carbon RLs but only the cool \HI\ component will be observed 
in self-absorption.

\vfill
\end{document}